# Integrated spatial multiplexing of heralded single photon sources

M. J. Collins[1], C. Xiong[1], I. H. Rey[2], T. D. Vo[3], J. He[1], S. Shahnia[1], C. Reardon[4], M. J. Steel[5], T. F. Krauss[4], A. S. Clark[1], and B. J. Eggleton[1,*]

[1]Centre for Ultrahigh bandwidth Devices for Optical Systems (CUDOS), the Institute of Photonics and Optical Science (IPOS), School of Physics, University of Sydney, NSW 2006, Australia.

[2]SUPA, School of Physics and Astronomy, University of St. Andrews, St. Andrews, Fife KY16 9SS, United Kingdom.

[3]Maritime Operations Division, Defence Science and Technology Organisation (DSTO), Department of Defence, P.O. Box 44, Pyrmont, NSW 2009, Australia.

[4]Department of Physics, University of York, York YO10 5DD, United Kingdom.

[5]CUDOS, MQ Photonics Research Centre, Department of Physics and Astronomy, Macquarie University, NSW 2019, Australia

*email: egg@physics.usyd.edu.au

The non-deterministic nature of photon sources is a key limitation for single photon quantum processors. Spatial multiplexing overcomes this by enhancing the heralded single photon yield without enhancing the output noise. Here the intrinsic statistical limit of an individual source is surpassed by spatially multiplexing two monolithic silicon correlated photon pair sources, demonstrating a 62.4% increase in the heralded single photon output without an increase in unwanted multi-pair generation. We further demonstrate the scalability of this scheme by multiplexing photons generated in two waveguides pumped via an integrated coupler with a 63.1% increase in the heralded photon rate. This demonstration paves the way for a scalable architecture for multiplexing many photon sources in a compact integrated platform and achieving efficient two photon interference, required at the core of optical quantum computing and quantum communication protocols.

It has been shown that the quantum nature of single photons could be used to create a quantum computer[1] and to communicate securely[2]. These initial proposals were formulated with the assumption that perfect photon sources would one day exist. Imperfect non-deterministic single photon sources have limited the complexity of demonstrations to date, despite significant work to simplify quantum operations and ease technical demands[3]. If the single photon output for multiple indistinguishable sources can be engineered to be more deterministic, this would create a new potential for scalable quantum technology based on photon-photon interference, including boson-sampling processors[4,5], long distance communication[6] and metrology[7].

Here we report the experimental demonstration of integrated, spatially multiplexed heralded single photon sources. A ceramic optical switch, made from ultra-low loss lead lanthanum zirconium titanate (PLZT)[8], was used to route photons from two monolithic silicon sources to a common output. Our results demonstrate that it is possible to increase the rate of useful single photons by multiplexing together many individual heralded single photon sources. We achieve 62.4% and 63.1% enhancement to the probability of generating a single photon from a pump pulse without affecting the noise level in two separate experiments, the first on two separately pumped sources and the second with two sources pumped through a common input. Scaling up to a larger number of devices will facilitate the creation of a new class of useful single photon sources.

Several physical systems can be used to generate single photons. Atom-like systems potentially offer on-demand single photons, but are difficult to implement on-chip and scale up. This is due to challenges in engineering atom-like quantum-dot sources to emit at the same frequency, large footprint bulk photon collection apparatus and the typical requirement for cryogenics[9]. Attenuated coherent light sources are frequently used as a source of non-deterministic single photons, but the photon statistics are fundamentally limited by a Poisson distribution. While easy to package, this Poissonian limit on the photon statistics cannot be improved upon, inevitably adding a multi-photon component to the output state. Heralded single photon sources are based on correlated photon pair generation. By detecting one photon from a pair, the existence and timing of the remaining photon is then known. This additional information is the basis of multiplexing where heralded single photons from multiple sources can be deterministically combined using active switching.

Heralded single photon sources are often based on photon-pairs generated in nonlinear optical media including, spontaneous four-wave mixing (SFWM) in $\chi^{(3)}$ waveguides[10-13] and spontaneous parametric down conversion in $\chi^{(2)}$ devices[14,15]. The number of photon pairs generated per pulse is governed by probabilistic statistics ranging from Poissonian to thermal depending on the filter bandwidth[16,17]. Multi-pair generation introduces noise photons, thus setting a fundamental relationship between the number of photon pairs generated per pulse and the noise performance of the source. For low-noise photon pair generation, sources must therefore operate in a regime where the probability of generating more than one pair per pulse is low, typically limiting the probability $\mu$ of generating one pair to be below 0.1 per pulse. We can introduce the collection efficiency $\eta$ as the probability a photon goes through the circuit to the detector; typically between $10^{-2}$ and $10^{-3}$ due to component loss and imperfect detection. The probability that one photon of a pair is detected is then $\mu\eta$ and for a heralded single photon or a two-photon coincidence is $\mu\eta^2$. The probability of heralded two-photon interference events is $\mu^2\eta^4$, meaning heralded two-photon operations are extremely uncommon. As heralded two-photon interference is at the core of many quantum gates, this is detrimental to the implementation of many algorithms[18-20].

Multiplexing many photon pair sources decouples the relationship between the desirable single photon output and multi-pair noise governing an individual source, allowing a higher single pair rate for a fixed multi-pair (noise) rate. Multiplexing can be implemented with both spatial[21-23] or temporal schemes[24]. The previous demonstration of efficient multiplexing used a free-space scheme[23] inherently sensitive to alignment stability. The advantages of an integrated photonic architecture, in contrast, are that many sources and components can be fabricated onto a single monolithic chip with proven stability[25]. While this strategy has been highly successful for tunable quantum processing[26], at present there has been no demonstration for scalable multiplexed single photon generation.

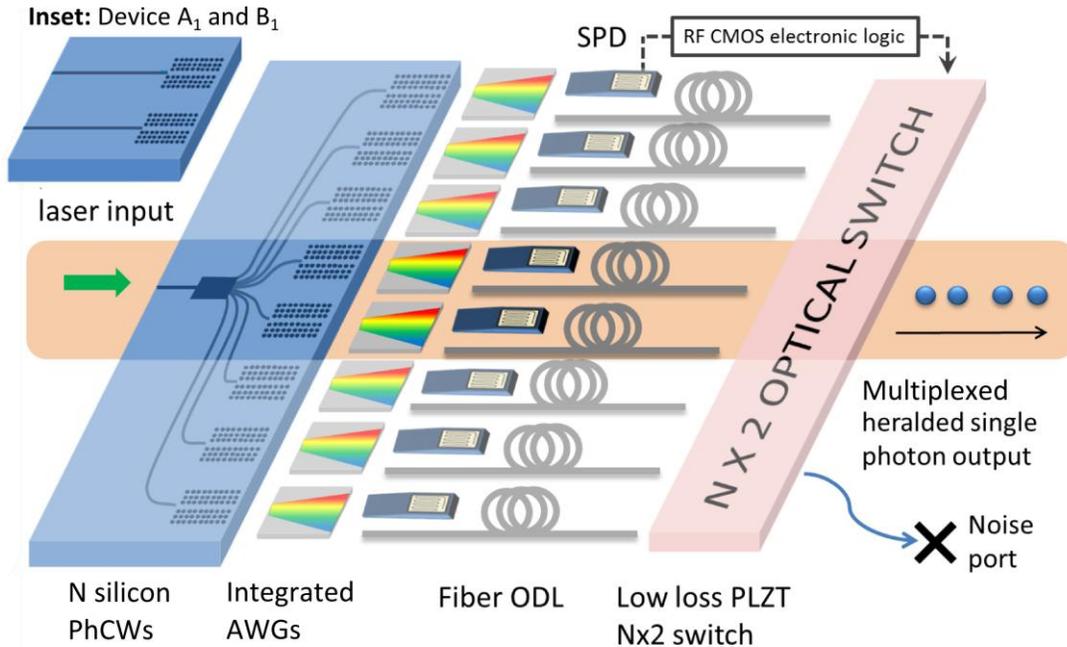

**Figure 1 | Spatial multiplexing scheme.** A single laser is coupled to a monolithic silicon chip to pump an ultra-compact array of silicon photonic crystal waveguides (PhCW). Photon pairs are generated in the PhCW region, wavelength separated by integrated arrayed waveguide gratings (AWGs) and the heralding photons detected using single photon detectors (SPD). The remaining photons go through a delay line, while a fast electronic logic gate sets the state of the PLZT switch. The selected heralded photon is then routed to the common fiber output. The orange box region represents the experimental setup for results presented in this work. Inset: Our first implementation used a device with two separate but monolithically fabricated PhCWs, designated $A_1$ and $B_1$.

Figure 1 shows a schematic of our proposed architecture for efficient spatial multiplexing of $N$ photon pair sources with the aim of building a more deterministic single photon source. Here, pump pulses are coupled to a silicon waveguide and split to a bank of $N$ nominally identical and monolithically integrated photonic crystal waveguides (PhCW), where photon-pairs are generated by SFWM. Slow-light engineering of the PhCW enhances the nonlinearity providing an ultra-compact device[27]. Furthermore, the position of the slow-light frequency window must be nominally identical in all PhCWs, to permit generation of indistinguishable photon pairs using a single pump laser. The two photons are separated using wavelength division components that can be realized, for example, using integrated arrayed waveguide gratings (AWGs)[28,29]. One photon from each pair is detected by an array of $N$ integrated silicon waveguide-based single photon detectors[30], while the other is buffered in an optical delay line (ODL)[31]. While the photon is delayed, an RF-logic gate triggered by a photon detection feeds forward a drive signal to a fiber coupled opto-ceramic switch[32], which actively routes single photons to a common output. For a net increase in the heralded photon rate in the case of only two sources, the switch insertion loss must be less than 3 dB[21]. The output is a multiplexed stream of indistinguishable single photons, given matched post generation spectral filtering.

Here we present spatial multiplexing of heralded single photons using the above architecture, for $N=2$, in two different configurations. The first consists of two PhCWs with the classical pump beam coupled separately to each waveguide on the chip, allowing for ease of balancing the coupled power. The second has, in addition, an on-chip 50:50 Y-coupler to divide the input pump laser between two PhCWs, demonstrating further integration.

**Results**

In the first demonstrations the two waveguides are labeled $A_1$ and $B_1$ and sketched in the inset of Fig. 1. The rate of pairs from each source was measured before the electro-optic switch and compared to the multiplexed rate measured at the common output. The signal-to-noise metric for probabilistic photon pair sources is the coincidence-to-accidental ratio (CAR). Fig. 2 illustrates the CAR for a range of measured heralded single photon rates for individual PhCWs $A_1$ and $B_1$, shown as red squares and green circles respectively. The characteristic curve for the measured CAR of an individual source is limited by multi-pair noise. The absolute rate is then estimated by taking into account component losses and detector efficiencies. We spatially multiplex the two sources by adding in an optical switch and electronics to route photons to a common output fiber. The multiplexing measurements were made using the same characterization setup as for the individual sources and the result is plotted in Fig. 2 as blue triangles. We fitted the data for the single sources using an analytical expression for the CAR including the measured values for detector dark counts and losses (see supplementary material). By measuring the transmission of the opto-ceramic switch channels (85.1% and 79.4%), we calculated the maximum expected enhancement to the CAR as a function of the heralded single photon rate (dashed line in Fig. 2, see Supplementary Material for more detail)[17]. It can be seen that the data is entirely consistent with this maximum expected enhancement and that the results for the multiplexing case lie well above the limit for a single source of this type. Indeed, to check the agreement, we fit the data to the predicted curve but allowing the switch transmission to vary. The best fit was obtained for a transmission that slightly exceeds the measured values, so that within experimental error we have obtained the maximum possible performance. We also show the theoretical case for a lossless switch and note that we are operating close to this regime. For a fixed CAR we achieved a 62.4% improvement in the detected single photon rate, equivalent to increasing the probability of generating a photon while maintaining a constant probability of multi-pair noise. Similarly, an improvement in the CAR was achieved for a fixed photon rate, which corresponds to a reduction in the multi-photon noise. The decoupling of the single photon and multi-photon rates, evident in the result shown in Fig. 2, is the essence of spatial multiplexing. To provide a clear demonstration, the measurement was made in the regime where multi-photon generation is the dominant noise source, at rates well above the dark-count limited CAR peak[13]. Note that other methods to enhance the single-photon rate, including increasing pump power or pump laser repetition rate[33], cannot be used to achieve the same decoupling, and the noise scales as for individual sources.

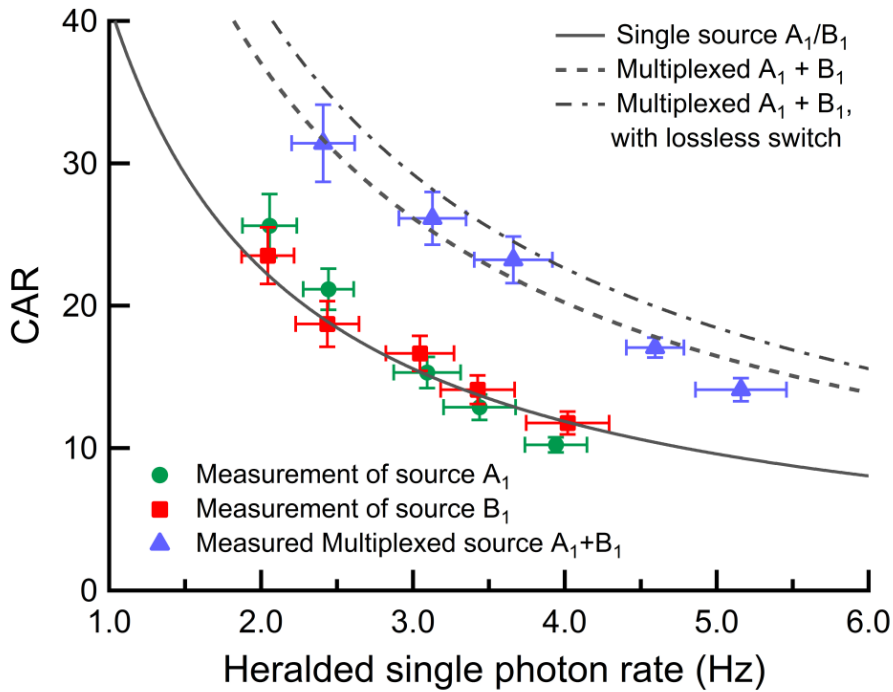

**Figure 2 | Spatial multiplexing result.** The coincidence-to-accidental ratio (CAR) from source $A_1$ (green circles), source $B_1$ (red squares) and after multiplexing (blue triangles) are plotted for a range of heralded single photon rates. All errors are calculated from Poissonian statistics.

The second experimental implementation demonstrates a primary building block for our proposed scheme and is shown schematically as the region inside the orange box in Fig. 1. The two PhCW devices are referred to as $A_2$ and $B_2$, shown in the scanning electron micrograph, Fig. 3A. The output heralded single photon rate from each waveguide was measured and plotted in Fig. 3B, shown as green triangles and red squares for $A_2$ and $B_2$ respectively. The multiplexing switch was then added and the multiplexed output measured for the same range of PhCW output powers, shown in Fig. 3B as blue triangles. We extract an enhancement to the heralded single photon rate of 63.1%, consistent with the result of the first measurement.

To verify that the output of our multiplexed source was indeed in the single photon regime, we measured the second-order correlation function $g^{(2)}(nT)$ at discrete delays using a Hanbury-Brown and Twiss (HBT) setup[34]. Here $n$ is an integer and $T$ is the period between pump laser pulses. A measured $g^{(2)}(0) < 1$ is expected for a non-classical light source and $g^{(2)}(0) < 0.5$ for a source approaching true single photon operation. A 50/50 coupler is required for a HBT experiment, and was added to the common output. An optical delay line was added to adjust the fine balance in optical path length (see Methods), with the requirement that the optical paths from each of the photon pair sources and the coupler be balanced. In this experiment, sources $A_2$ and $B_2$ both contribute to the measured photon statistics used to calculate $g^{(2)}(nT)$, which is plotted in Fig. 3C. The measured $g^{(2)}(0) = 0.17 \pm 0.03$ confirms our spatially multiplexed source is operating in the single photon regime. As a check of this result, we measured $g^{(2)}(nT)$ for $n = \pm 1$, delaying the electronic trigger pulses from the detector at one output of the 50/50 coupler by one pulse period with respect to the other and measuring the correlation. For these correlations between successive pulses, we expect the photon statistics to return to a Poissonian distribution with $g^{(2)}(T) = 1$. The measured $g^{(2)}(+T) = 1.06 \pm 0.3$ and $g^{(2)}(-T) = 0.96 \pm 0.3$ are consistent with this expectation.

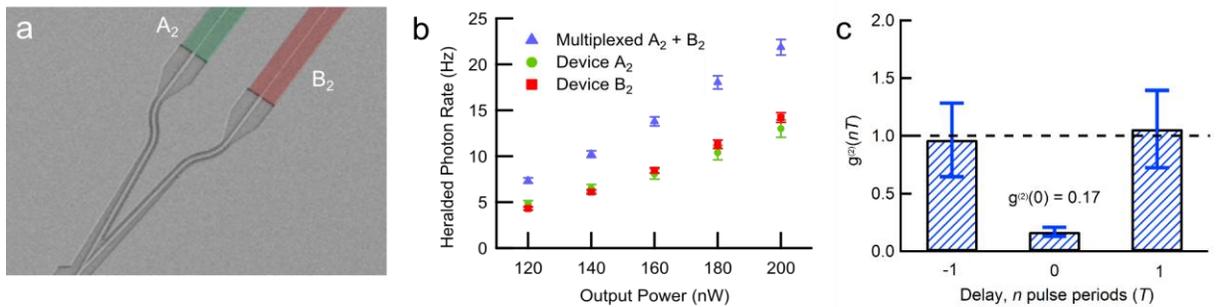

**Figure 3 | On-chip Y-split coupler.** (a) A scanning electron microscope image of an integrated waveguide based Y-split geometry 50/50 coupler, splitting incoming pump pulses between two photonic crystal waveguides $A_2$ and $B_2$ shown in green and red false colour respectively. (b) Heralded single photon rates from source $A_2$ (green circles), source $B_2$ (red squares) and the multiplexed (blue triangles) rate are plotted for a range of input powers. (c) The $g^{(2)}(nT)$ correlation function is measured for the multiplex output, a value of $g^{(2)}(0) = 0.17$, markedly less than 0.5, confirms the source is operating in the single photon regime. All errors are calculated from Poissonian statistics.

**Discussion**

If we were to extend this demonstration, for example scaling to eight sources requiring the realistically achievable number of nine detectors, this would provide an enhancement of the heralded single photon rate by a factor of five[35,36]. This corresponds to a greater than 20-fold increase in the two-photon interference events required for the operation of, for example, a quantum relay[37] or a controlled-NOT gate[25,38]. PhCWs are ideal sources for this scheme due to their broadband slow-light enhancement that enables compact device design even with a spectrally broad pulsed pump. The efficiency of the heralding detectors is currently limited to between 10% and 25% when operating at 1550 nm. In a multiplexed source, any improvement to the detectors carries over favorably to the single photon output rate. For example doubling the detection efficiency for a source with $N$ multiplexed waveguides would result in a $2^N$ increase in the heralded single photon rate. The linear losses associated with the waveguide and filtering components will likely be reduced in step with advances in fabrication technology, further increasing the single photon rate.

This demonstration provides a road-map for the creation of fully integrated near-deterministic heralded single photon sources. Our results show that integrated spatial multiplexing can be implemented efficiently to overcome the non-deterministic nature of single photon sources currently limiting applications requiring many single photons. We achieved an enhancement to the probability of generating a single photon of 62.4% and 63.1%, breaking the intrinsic limit of a single source by decoupling the single and multi-pair noise probabilities. Our building-block demonstration establishes the feasibility of scalable multiplexed sources, which promise a new domain of fundamental investigations in quantum mechanics including high-photon number entanglement, complex quantum computation schemes and highly efficient quantum communication. Multiplexing will continually benefit from technological development in the areas of ultra-high efficiency detectors, low loss integrated optics and precision fabrication leading to more impressive multi-qubit (>10) demonstrations using a stable source.

**Methods**

**Device fabrication.** The silicon photonic crystal waveguide (PhCW) devices were fabricated from a silicon-on-insulator wafer, with 220 nm thick Si layer above a 2 μm thick layer of silica. The photonic crystal was patterned using electron beam lithography and reactive ion etching to create a triangular lattice of holes. The waveguides were made by introducing a row defect. The photonic crystal region was then undercut, by etching away the silica substrate in that region, suspending the PhCWs in air. The two rows of holes adjacent to the waveguide were laterally shifted to engineer the dispersion[27] such that the group index was approximately 30 across a

~15 nm bandwidth and centered at a wavelength of 1559 nm. The effective nonlinearity ($\gamma_{eff} = (n_g/n_0)\gamma$) was approximately 4000 W$^{-1}$m$^{-1}$ where the slow-down factor, calculated as the ratio of the group index $n_g$ to the native refractive index of the material $n_0$ is included. The PhCW region had a linear propagation loss of ~50 dB.cm$^{-1}$. The waveguides were 196 μm long with inverse tapers and SU8 polymer cladding for improved coupling by mode matching to input-output lensed fibers. The first device measured had many separate PhCWs with individual input and output coupling. The second sample included Y-split couplers, fabricated using silicon nanowires preceding the pairs of PhCWs. The PhCW regions were fabricated as close as possible to each other, to avoid variations in the Si layer thickness across the surface of the wafer, minimizing any differences in the dispersion and group index.

**Experimental methods for spatial multiplexing.** In the first experiment ~7 ps laser pulses were separated off-chip with a 50:50 fiber-based directional-coupler and coupled to on-chip polymer access waveguides using a lensed fiber. For the second experiment a Y-split coupler was integrated on to the same chip as the PhCWs and the pump coupled to the device via a single lensed fiber and input polymer waveguide. The pulses were thus divided and then coupled into the two PhCWs which had nominally identical dispersion properties. In the PhCW regions, correlated photon pairs were generated via SFWM. Each pair was coupled out of the device, again using inverse tapers and lensed fibers to minimize loss. The photons from each pair were separated into two single mode optical fibers using a pair of arrayed waveguide gratings (AWG) with a photon channel detuning of 300 GHz from the pump channel. The higher energy photons, labeled the signal photons, were detected using superconducting single photon detectors (SSPD). These detection events provided the heralding electrical pulses that triggered the optical switch. Depending on whether source A or B generated a pair, the optically transparent ceramic[8] switch routed the remaining lower energy photon from each pair, referred to as the idler, to the common output fiber where they were detected using a third SSPD. All counts were then analysed for coincident detection using a time interval analyser.

**Measuring the second-order correlation function.** We measured the second-order correlation function $g^{(2)}(nT)$ after multiplexing the photons from two PhCWs using a Hanbury-Brown and Twiss setup. The setup was modified by adding a 50:50 directional-coupler to the common output and a fourth detector, noting $n$ is an integer and $T$ is the time between each successful pump pulse, together $nT$ is the difference in arrival time to the detector between photons exiting the two outputs of the coupler. At zero time delay between the detectors a $g^{(2)}(0) < 1$ is expected for a non-classical light source and $g^{(2)}(0) < 0.5$ for a source approaching single photon operation. The measured $g^{(2)}(0) = 0.17 \pm 0.03$ confirms the spatially multiplexed source is operating in the single-photon regime. We verified the validity of our measurement by determining $g^{(2)}(\pm 1)$ for neighboring pulses which were found to be unity within error, consistent with the Poissonian statistics expected when measuring distinguishable photons[16,17].

**References**

1. E. Knill, R. Laflamme, G. Milburn, A scheme for efficient quantum computation with linear optics. *Nature* **409**, 46-52 (2001).
2. H. J. Kimble, Quantum internet. *Nature* **453**, 1023-1030 (2008).

3. E. Martin-Lopez *et al.,* Experimental realization of Shor's algorithm using qubit recycling. *Nat. Phot*onics **6**, 773-776 (2012).

4. J. B. Spring *et al.,* Boson sampling on a photonic chip. *Science* **339**, 798-801 (2013).

5. M. A. Broome *et al.,* Photonic boson sampling in a tunable circuit. *Science* **339**, 794-798 (2013).

6. N. Gisin, R. Thew, Quantum communication. *Nat. Photonics* **1**, 165-171 (2007).

7. T. Nagata, R. Okamoto, J. L. O'Brien, K. Sasaki, S. Takeuchi, Beating the standard quantum limit with four-entangled photons. *Science* **316**, 726-729 (2007).

8. H. Jiang *et al.,* Transparent electro-optic ceramics and devices. *Proc. SPIE* **5644***, Optoelectronic Devices and Integration,* 380 (2005).

9. S. V. Polyakov *et al.,* Coalesence of single photons emitted by disparate single-photons sources: the example of InAs quantum dots and parametric down-conversion sources. *Phys. Rev. Lett.* **107**, 157402 (2011).

10. J. E. Sharping *et al.,* Generation of correlated photons in nanoscale silicon waveguides. *Opt. Express* **14**, 12388-12393 (2006).

11. H. Takesue *et al.,* Entanglement generation using silicon wire waveguide. Appl. Phys. Lett. **91**, 201108 (2007).

12. S. Clemmen *et al.,* Continuous wave photon pair generation in silicon-on-insulator waveguides and ring resonators. *Opt. Express* **17**, 16558-16657 (2009).

13. C. Xiong *et al.,* Characterisitics of correlated photon pairs generated in ultracompact silicon slow-light photonic crystal waveguides. *IEEE J. Sel. Top. Quantum Electron.* **18**, 1676-1683 (2012).

14. J. Chen, A. J. Pearlman, A. Ling, J. Fan, A. Migdall, A versatile waveguide source of photon. *Opt. Express*, **17**, 6727-6740 (2006).

15. M. Hunault, H. Takesue, O. Tadanaga, Y. Nishida, M. Asobe, Generation of time-bin entangled photon pairs by cascaded second-order nonlinearity in a single periodically poled LiNbO$_3$ waveguide. *Opt. Lett.* **14**, 12388-12391 (2010).

16. H. De Reidematten, *et al.,* Two independent photon pairs versus four-photon entangle state in parametric down conversion. *J. Mod. Opt.* **51**, 1637-1649 (2004).

17. H. Takesue, K. Shimizu, Effects of multiple pairs on visibility measurements of entangled photons generated by spontaneous parametric processes. *Opt. Commun.* **283**, 276-287 (2010).

18. M. J. Collins *et al.,* Low Raman-noise correlated photon-pair generation in a dispersion-engineered chalcogenide As$_2$S$_3$ planar waveguide. *Opt. Lett.* **37**, 3393-3395 (2012).

19. M. Davanço *et al.,* Telecommunications-band heralded single photons from a silicon nanphotonic chip. *Appl. Phys. Lett.* **100**, 261104 (2012).

20. W. C. Jiang, X. Lu, J. Zhang, O. Painter, Q. Lin, A silicon-chip source of bright photon-pair comb. (Available at http://arxiv.org/abs/1210.4455v1).

21. A. L. Migdall, D. Branning, S. Castelletto, Tailoring single-photon and multiphoton probabilities of a single-photon on-demand source. *Phys. Rev. A*. **66**, 053805 (2002).

22. J. H. Shapiro, F. N. C. Wong, On-demand single-photon generation using a modular array of parametric downconverters with electro-optic polarization controls. *Opt. Lett.* **32**, 2698-2700 (2007).

23. X.-S. Ma, S. Zotter, J. Kofler, T. Jennewein, A. Zeilinger, Experimental generation of single photons via active multiplexing. *Phys. Rev. A* **83**, 043814 (2011).

24. J. Mower, D. Englund, Efficient generation of single and entangled photons on a silicon photonic integrated chip. *Phys. Rev. A* **84**, 052326 (2011).

25. A. Politi, M. J. Cryan, J. G. Rarity, S. Yu, J. L. O'Brien, Silica-on-silicon waveguide quantum circuits. *Science* **320**, 646-649 (2008).

26. P. J. Shadbolt *et al.,* Generating, manipulating, and measuring entanglement and mixture with a reconfigurable photonic circuit. *Nat. Photonics* **6**, 45-49 (2012).

27. J. Li, T. P. White, L. O'Faolain, A. Gomez-Iglesias, T. F. Krauss, Systematic design of flat band slow light in photonic crystal waveguide. *Opt. Express* **16**, 6227-6232 (2008).

28. T. Tsuchizawa *et al.,* Microphotonics devices based on silicon microfabrication technology. *IEEE J. Sel. Top. Quantum Electron.* **11**, 232-240 (2005).

29. D. Dai *et al.,* Low-loss $Si_3N_4$ arrayed-waveguide grating (de)multiplexer using nano-core optical waveguides. *Opt. Express* **19**, 14130-14136 (2011).

30. J. P. Sprengers *et al.*, Waveguide superconducting single-photon detectors for integrated quantum photonic circuits. *Appl. Phys. Lett.* **99**, 181110 (2011).

31. G. D. Marshall *et al.,* Laser written waveguide photonic quantum circuits. *Opt. Express* **17**, 12546-12554 (2009).

32. K. Nashimoto, "Nano-second speed PLZT waveguide switches and filters" (EpiPhotonics, San Jose, CA, USA, 2011).

33. M. A. Broome, M. P. Almeida, A. Fedrizzi, A. G. White, Reducing multi-photon rates in pulsed down-conversion by temporal multiplexing. *Opt. Express* **19**, 545-550 (2011).

34. R. Hanbury Brown, R. Q. Twiss, Correlation between photons in two coherent beams of light. *Nature* **177**, 27-29 (1956).

35. E. Jeffrey, N. A. Peters, P. G. Kwiat, Towards a periodic deterministic source of arbitrary single-photons. *New J. Phys.* **6**, 100 (2004).

36. A. Christ, C. Silberhorn, Limits on the deterministic creation of pure single-photon states using parametric down-conversion. *Phys. Rev. A* **85**, 023829 (2012).

37. A. Martin, O. Alibart, M. P. De Micheli, D. B. Ostrowsky, S. Tanzilli, A quantum relay chip based on telecommunication integrated optics technology. *New J. Phys.* **14**, 025002 (2012).

38. A. S. Clark, J. Fulconis, W. J. Wadsworth, J. G. Rarity, J. L. O'Brien, All-optical-fiber polarization-based quantum logic gate. *Phys. Rev. A* **79**, 030303 (2009).

**Acknowledgments**

This work was supported in part by the Centre of Excellence (CUDOS, project number CE110001018), Laureate Fellowship (FL120100029) and Discovery Early Career Researcher Award (DE130101148 and DE120100226) programs of the Australian Research Council (ARC) and EPSRC UK Silicon Photonics (Grant reference EP/F001428/1).